\documentclass[10pt,final,journal,twocolumn]{IEEEtran}
\ifCLASSINFOpdf
\else
\fi
\usepackage{cite}
\usepackage[cmex10]{amsmath}
\usepackage{amsthm}
\usepackage{algorithmic}
\usepackage{stfloats}
\usepackage{multicol,multienum}
\usepackage[dvips]{graphicx}
\begin{document}

\title{Load-aware Dynamic Spectrum Access for Small Cell Networks: A Graphical Game Approach }

\author{Yuhua~Xu,~\IEEEmembership{Member,~IEEE,}
        Chenggui Wang,
        Junhong Chen,
        Jinlong~Wang~\IEEEmembership{Senior Member,~IEEE,}
        Yitao Xu,
        Qihui~Wu,~\IEEEmembership{Senior Member,~IEEE,}
        and Alagan~Anpalagan,~\IEEEmembership{Senior Member,~IEEE}
%
%
}

%
%

\IEEEpeerreviewmaketitle
\maketitle

\begin{abstract}
In this letter, we investigate the problem of dynamic spectrum access for small cell networks, using a graphical game approach. Compared with existing studies, we  take the features of different cell loads and local interference relationship into account.  It is proved that the formulated spectrum access game is an exact potential game with the aggregate interference level as the potential function, and Nash equilibrium (NE) of the game corresponds to the global or local optima of the original optimization problem. A lower bound of the achievable aggregate interference level is rigorously derived. Finally, we propose an autonomous best response learning algorithm to converge towards its NE. It is shown that the proposed game-theoretic solution converges rapidly and its achievable performance is close to the optimum solution.
\end{abstract}

\begin{IEEEkeywords}
  5G networks, small cell networks, dynamic spectrum access, potential game.
\end{IEEEkeywords}

%
\IEEEpeerreviewmaketitle

\section{Introduction}
\IEEEPARstart {S}{mall} cell is  an enabling technology for 5G networks, since it has been regarded as the most promising approach for providing  a thousand-fold mobile traffic over the next decade \cite{small_cell1}. Technically, the use of very dense and low-power small cells exploits the following two fundamental effects \cite{Cloud_5G}: i) the decreasing distance between the  base station and the user leads to higher transmission rates, and ii) the spectrum is more efficiently exploited due to the improved spectrum spatial reuse ratio. As the network becomes denser, temp-spatial variations of mobile traffics can be observed and the small cells are usually deployed randomly and dynamically \cite{self_organizing}. As a result,  traditional centralized optimization approaches, e.g., the  graph coloring algorithm \cite{Graph_coloring}, can not be applied in practice. To overcome this shortage, there are some distributed spectrum access approaches using,  e.g., sensing-based access approach \cite{distributed_spectrum_SC}, utility-based learning approach  \cite{Utility_learning}, reinforcement-learning based self-organizing scheme \cite{Game_SmallCell_1}, coalitional game based scheme \cite{Game_SmallCell_2}, evolutionary game based scheme \cite{Game_SmallCell_3} and hierarchical dynamic game approach  \cite{Game_SmallCell_4}.

However, there are two limitations in existing distributed approaches: i) the fact that the small cells have different loads was not addressed, i.e., most existing work assumed that there is only one mobile user in each small cell, and ii) the feature of local interference due to the low transmission power, e.g., the transmission of a small cell only directly affects its nearby cells, was not considered. In this letter, we consider load-aware dynamic spectrum access for small cell networks, taking into account different cell loads and local interference relationship.


We consider a sensing-based autonomous spectrum access mechanism, i.e., a small cell transmits on the channels which are detected idle  \cite{distributed_spectrum_SC}. In such scenarios, it is desirable to decrease the number of neighboring cells choosing the same channel. We first define a new optimization metric to capture the interference among the small cells. Then, we formulate the spectrum access problem as a graphical game and propose a self-organizing distributed spectrum access algorithm. To summarize, the contributions of this letter are:
\begin{enumerate}
  \item We formulate the spectrum access problem for the small cells as a graphical game, taking the inherent features of  different cell loads and local interference relationship  into account. It is proved that it is an exact potential game with the aggregate interference level as the potential function; furthermore, the Nash equilibrium (NE) of the game corresponds to the global or local optima of the original problem. Also, a lower bound of the  aggregate interference level is rigorously derived.
  \item We propose an autonomous best response (BR) algorithm to converge towards NE of the game. Compared with the standard BR algorithm, the proposed  algorithm converges rapidly and is scalable when the number of small cells becomes large. Simulation results show that its performance is very close to the global optimum.
\end{enumerate}

Note that game-based spectrum access approaches have been extensively used in the literature \cite{Utility_learning,Game_SmallCell_1,Game_SmallCell_2,Game_SmallCell_3, Game_SmallCell_4, La12,Buzzi12}. In methodology, the differences and new challenges in this work are: i) in existing work, it is assumed that there is only one serving mobile user in each cell and the task is to choose an operational channel. When different cell loads are considered, each cell generally needs multiple channels rather than one channel. However, existing game design and analysis with singleton action selection can not be applied. ii) we consider a graphical game model, i.e.,  the direct interaction only exists between neighboring users, and hence is significantly different from previous global interactive game models, i.e., the interaction emerges among all  users, iii) we define a new  metric to capture the interference relationship among neighboring small cells.


The rest of this letter is organized as follows. In Section II,  the system model and  problem formulation are presented. In Section III,  the graphical game model is formulated and analyzed, and an autonomous best response learning algorithm is proposed to achieve its NE. Finally, simulation results and discussion are presented in Section IV and conclusion is drawn in Section V.

\vspace{-0.1in}
\section{System Model and Problem Formulation}
Consider a small cell network consisting of $N$  small cell access points (SAPs) and each SAP serves several mobile users (MUs).
It is assumed that the small cells and the macro-cell operate on orthogonal channels, and hence the main  optimization objective is eliminating mutual interference among the small cells. Note that this assumption has been extensively used in previous work \cite{Graph_coloring,Utility_learning,Estimation_load,adaptive_allocation,Gateway}. Also,  it is in line with 3GPP \cite{3GPP} and particularly represents the scenarios in the LTE-U network \cite{LTE_U}, which is an active research topic.

There are $M$ channels available for the SAPs.
Denote the SAP set as $\mathcal{N}$, i.e., $\mathcal{N}=\{1,\dots,N\}$, and the available channel set as $\mathcal{M}$, i.e., $\mathcal{M}=\{1,\dots,M\}$. It is shown that as the small cells become denser in 5G networks, the more spatial load fluctuation is observed by each SAP \cite{Cloud_5G}. To capture such a  fluctuation, it is assumed that  each SAP chooses $K_n$ channels for data transmission of the MUs.  The number  $K_n$ can be regarded as the  load of each SAP, which is jointly determined by the number of active MUs and their traffic demands\footnote{Furthermore, since the users in the small cells are always random and dynamic, it is not reasonable to allocate spectrum resources based on the instantaneous network state; instead, it is preferable to allocate spectrum resources according to their loads in a relatively longer decision period.}. Similar to previous work \cite{Buzzi12,La12,Utility_learning}, we focus on the spectrum access problem and do not consider the problem of optimizing the required number of channels of each SAP. In practice, some simple but efficient approaches, e.g., the  one proposed in \cite{Estimation_load}, can be applied to estimate the cell load.

Due to the spatial distribution and lower transmission power of SAPs, the transmission of a small cell only directly affects the neighboring small cells \cite{interference_graph1,Graph_coloring,Estimation_load,Gateway}. To characterize the interference relationship among the small cells, the following interference graph is introduced. Specifically, if the distance $d_{ij}$ between  SAP $i$ and $j$ is lower than a threshold $d_0$,  then they interfere with each other when transmitting on the same channel. Therefore, the potential interference relationship can be captured by an interference graph $\mathcal{G}=\{V, E\}$, where $V$ is the vertex set (the SAP set) and $E$ is the edge set, i.e., $V=\{1,\dots,N\}$ and $E=\{(i,j)|i\in\mathcal{N}, j\in\mathcal{N}, d_{ij}<d_0\}$. For presentation, denote the neighboring SAP set of SAP $n$ as $\mathcal{J}_n$, i.e., $\mathcal{J}_n=\{j\in\mathcal{N}:d_{nj}<d_0\}$.

If two or more neighboring small cells choose the same channel, mutual interference may occur. Thus, in order to mitigate interference among the small cells, it is desirable to allocate non-overlapping channels for them as soon as possible.
Denote the choice of channels by SAP $n$ as $a_n=\{c_1,c_2,\ldots,c_{K_n}\}, c_i \in \mathcal{M}, \forall 1 \le i \le K_n$. Note that $a_n$ is a $K_n$-combination of $\mathcal{M}$ and the number of all possible chosen channel profiles of player $n$
is $C^{K_n}_M=\frac{M(M-1)\ldots(M-K_n+1)}{K_n(K_n-1)\ldots1}$. Motivated by the graph coloring for spectrum allocation problems \cite{Graph_coloring}, we define the experienced interference level as following:

\begin{equation}
 s_n= \sum_{j\in\mathcal{J}_n} \sum_{e\in a_n}  \sum_{f\in a_j}\delta(e,f),
\label{eq:interference_level}
\end{equation}
where $\delta(e,f)$ is the following {indicator function}:
\begin{equation}
\label{eq:delta_function}
\delta(e,f) = \left\{ \begin{array}{l}
 1,\;\;e=f   \\
 0,\;\;e \ne f. \\
 \end{array} \right.
\end{equation}
That is, if two selected channels $e$ and $f$ are the same, then the indication function takes one; otherwise, it takes zero.

 The rationale behind the  experienced interference level is briefly explained as follows:
in autonomous small cell networks, a small cell transmits only when the received energy on the dedicated channel is below a threshold. This is similar to the carrier sense multiple access and has been regarded as a proposing approach for cognitive small cell networks \cite{distributed_spectrum_SC} and LTE-U small cells \cite{LTE_U}. Therefore, decreasing the number of interfering cells would increase the achievable throughput.

\begin{figure}[!tb]
\centering
\includegraphics[width=2.4in]{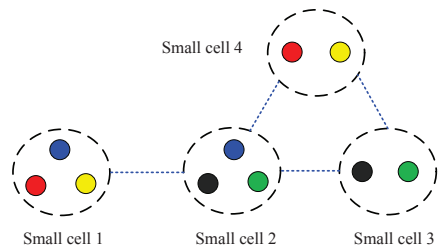}
\caption{An illustration for the considered interference model, in which different colors represent different channels.
To reduce the interference in the network, i) for intra-cell spectrum access, it is mandatory to allocate different channels for the users in the same cell, ii) for inter-cell spectrum access, the number of overlapping channels should be minimized. According to (\ref{eq:interference_level}), the interference levels of the cells are $s_1=1, s_2=3, s_3=2, s_4=0$. }
\label{fig:interference}
\end{figure}

Note that $s_n$  is the number of channels also chosen by the neighboring SAPs. For an individual SAP $n$, the interference level $s_n$ should be minimized. From a network-centric perspective, the aggregate interference level of all the SAPs, i.e., $\sum\nolimits_{n\in \mathcal{N}}s_n$ should be minimized. The considered interference model is illustratively depicted in Fig. \ref{fig:interference}. Thus, we formulate the problem of load-aware spectrum allocation for cognitive small cell networks as follows:

\begin{equation}
\label{eq:social_optimum}
P1:\;\;\;\; \min \;  \sum_{n\in \mathcal{N}}s_n.
\end{equation}

It is noted that the definition of  the  interference model is different from that of traditional PHY-layer interference. Here, the interference level is used to characterize the mutual influence among neighboring SAPs from a higher-level view.  Such an interference model has also been applied for single channel selection in opportunistic spectrum access networks \cite{MAC_Interference1,MAC_Interference2,MAC_Interference3}. In comparison, this work extends previous single channel selection to load-aware multiple channel access. With the allocated channels, the small cell can perform power control to further reduce the mutual interference among different cells. However, this problem is beyond the scope of this letter.


\vspace{-0.1in}
\section{Graphical Game Model and Distributed Learning Algorithm}

To implement self-organizing and distributed spectrum access, we formulate a graphical game model to address the local interference relationship among the cells. The game is proved to be an exact potential game, and then a distributed learning algorithm is proposed to achieve its Nash equilibria.



\vspace{-0.1in}
\subsection{Graphical Game for Dynamic Spectrum Access}
Formally, the spectrum access game is denoted as $\mathcal{F}=[\mathcal{N}, \mathcal{G},{\{{\mathcal{A}_n}\}_{n\in\mathcal{N}}},{\{{u_n}\}_{n\in\mathcal{N}}}]$, where  $\mathcal{N} = \{1,\ldots,N\}$ is a set of players (small cells), $\mathcal{G}$ is the potential interference graph among the players, $\mathcal{A}_n=\{1,\ldots,M\}$ is a set of the available actions (channels) for each player $n$,  and $u_n$ is the utility function of player $n$.
Due to the limited interference range, the utility function can be expressed as $u_n(a_n,a_{\mathcal{J}_n})$, where $a_n$ is the action of player $n$ and  $a_{\mathcal{J}_n}$ is the action profile of the neighboring players of $n$. Thus, the formulated spectrum access game belongs to \emph{graphical game}. As discussed before, each small cell prefers  a lower interference level, which motivates us to define  the utility function as follows:
\begin{equation}
\label{eq:utility_function}
u_n(a_n,a_{\mathcal{J}_n})=-s_n,
\end{equation}
where $s_n$ is the experienced interference level of player $n$, as characterized by (\ref{eq:interference_level}). The players in the game are selfish and rational to maximize their individual utilities, i.e.,
\begin{equation}
\label{eq:game_model}
(\mathcal{F}): \;\;\;\;\;\;\;\; \mathop {\max }\limits_{a_n \in A_n} u_n(a_n,a_{\mathcal{J}_{n}}),\forall n \in \mathcal{N}.
\end{equation}

To analyze the properties of the formulated spectrum access game, we first present the following definitions.
\\
\textbf{Definition 1 (Nash equilibrium \cite{Monderer96})}. An action  profile $a^*=(a^*_1,\ldots,a^*_N)$ is a pure strategy  Nash equilibrium (NE)  if and only if no player can improve its utility  by deviating unilaterally, i.e.,
\begin{equation}
\label{eq:NE_definition}
 {u_n}({a^*_n},{a^*_{\mathcal{J}_n}}) \ge  {u_n}(a_n,{a^*_{\mathcal{J}_n}}), \forall n \in \mathcal{N}, \forall a_n \in \mathcal{A}_n, a_n\ne a^*_n
\end{equation}
\textbf{Definition 2 (Exact potential game \cite{Monderer96})}. A game is an exact potential game (EPG) if there exists an ordinal potential function $\phi: {{A}_1} \times  \cdots  \times {{A}_N} \to R$ such that for all $n \in \mathcal{N}$, all $a_n \in \mathcal{A}_n$, and $a'_n \in \mathcal{A}_n$, the following holds:
 \begin{equation}
 \label{eq:EPG_definition}
  \begin{array}{l}
    u_n(a_n,a_{\mathcal{J}_n})-u_n(a'_n,a_{\mathcal{J}_n}) = \phi(a_n,a_{\mathcal{J}_n})-\phi(a'_n,a_{\mathcal{J}_n})
  \end{array}
 \end{equation}
In other words, the change in the utility function caused by the unilateral action change of an arbitrary player is exactly the same with that in the potential function. It is known that EPG admits the following two promising features: (i) every EPG has at least one pure strategy NE, and (ii) an action profile that maximizes the potential function is also a NE.


\newtheorem{theorem}{Theorem}
\begin{theorem}
\label{tm:potential_game}
 The formulated spectrum access game $\mathcal{F}$ is an EPG, which has at least one pure strategy Nash equilibrium. In addition, the global optima of problem $P1$ are pure strategy Nash equilibria of $\mathcal{F}$.
\end{theorem}

\begin{IEEEproof}
To prove this theorem, we first construct the following potential function:
\begin{equation}
\label{eq:potential}
\Phi  ({a_n},{a_{ - n}}) =- \frac{1}{2} \sum \limits_{n\in \mathcal{N}} s_n(a_1,\ldots,a_N),
\end{equation}
where $s_n$ is characterized by (\ref{eq:interference_level}).

Recalling that the chosen channels of player $n$ is denoted as $a_n=\{c_1,c_2,\ldots,c_{K_n}\}$, define ${\mathcal{I}_n}({c_i},{a_{{J_n}}})$ as the set of neighboring players choosing a channel $c_i$,  $1 \le i \le K_n$, i.e.,
\begin{equation}
\label{eq:Interfering user set}
{\mathcal{I}_n(c_i,a_{J_{n}})} = \{j\in \mathcal{J}_n: {c_i} \in {a_j} \},
\end{equation}
where $\mathcal{J}_{n}$ is the neighbor set of player $n$. Then,  we denote
\begin{equation}
\label{eq:individual_collision_level2}
    s_n(c_i,a_{J_{n}})=|{\mathcal{I}_n(c_i,a_{J_{n}})}|
\end{equation}
as the experienced interference level on channel $c_i$, where $|A|$ is the cardinality of set $A$, i.e., the number of elements in $|A|$. Accordingly, the aggregate experienced interference level of player $n$ is also given by:
\begin{equation}
\label{eq:aggregate_level2}
    s_n(a_n,a_{J_{n}})=\sum \limits_{e \in a_n} s_n (e,a_{J_{n}})
\end{equation}

Now, suppose that an arbitrary player $n$ unilaterally changes its channel selection from $a_n=\{c_1,c_2,\ldots,c_{K_n}\}$ to $a^*_n=\{c^*_1,c^*_2,\ldots,c^*_{K_n}\}$. For presentation, we classify the channels into the following three sets:
\begin{itemize}
  \item $\mathcal{C}_0=a_n \cap a^*_n$. That is, the channels in set $\mathcal{C}_0$ are chosen by player $n$ both before and after its unilateral action change. Note that $\mathcal{C}_0$ may be a null set.
  \item $\mathcal{C}_1=a_n \backslash \{a_n \cap a^*_n\}$, where $A \backslash B$ means that $B$ is excluded from $A$. That is, the channels in $\mathcal{C}_1$ are only chosen by player $n$ before its unilateral action change.
  \item $\mathcal{C}_2=a^*_n \backslash \{a_n \cap a^*_n\}$. That is, the channels in $\mathcal{C}_2$ are only chosen by player $n$ after its unilateral action change.
\end{itemize}

From the above classification,  the change in utility function of player $n$ caused by  its action unilateral action change
is given by:
\begin{equation}
\label{eq:utility_change}
{u_n}(a_n^*,{a_{{J_n}}}) - {u_n}({a_n},{a_{{J_n}}})=
\sum \limits_{e \in \mathcal{C}_1}  s_n (e,a_{J_{n}})- \sum \limits_{e \in \mathcal{C}_2} s_n (e,a_{J_{n}})
\end{equation}

Also,   the change in the potential function caused by the unilateral change of player $n$ is as follows:
\begin{equation}
\label{eq:potential_change}
\begin{array}{l}
 \Phi  (a_n^*,{a_{ - n}}) - \Phi ({a_n},{a_{ - n}}) \\
  = \frac{1}{2} \Big\{ {u_n}(a_n^*,{a_{{J_n}}}) - {u_n}({a_n},{a_{{J_n}}}) \\
 \;\;\;\; + \sum\limits_{k \in \mathcal{D}_1} \big\{{u_k}(a_k,{a^*_{{J_k}}}) - {u_k}({a_k},{a_{{J_k}}})\big\}  \\
 \;\;\;\; + \sum\limits_{k \in \mathcal{D}_2} \big\{{u_k}(a_k,{a^*_{{J_k}}}) - {u_k}({a_k},{a_{{J_k}}})\big\}  \\
\;\;\;\;+ \sum\limits_{k \in \mathcal{D}_3, k \ne n } \big\{{u_k}(a_k,{a^*_{{J_k}}}) - {u_k}({a_k},{a_{{J_k}}})\big\} \Big\}, \\
 \end{array}
\end{equation}
where $\mathcal{D}_1=\underset {e \in \mathcal{C}_1} \cup {\mathcal{I}_n(e,a_{J_{n}})} $,
$\mathcal{D}_2=\underset {e \in \mathcal{C}_2} \cup {\mathcal{I}_n(e,a_{J_{n}})} $,   $\mathcal{D}_3=\mathcal{N} \backslash \{\mathcal{D}_1 \cup \mathcal{D}_2\}$, and ${u_k}(a_k,{a^*_{{J_k}}})$ is the utility function of player $k$ after $n$'s unilateral action change. Note that player $n$ belongs to the neighboring player set of player $k$, i.e., $n\in \mathcal{J}_k$.
Since the action of player $n$ only affects its neighboring players, the following equations hold:
\begin{equation}
\label{eq:potential_change1}
{{u_n}(a_k,{a^*_{{J_k}}}) - {u_n}({a_k},{a_{{J_k}}})}  = 1,\forall k \in \mathcal{D}_1
\end{equation}
\begin{equation}
\label{eq:potential_change2}
{{u_n}(a_k,{a^*_{{J_k}}}) - {u_n}({a_k},{a_{{J_k}}})}  = -1,\forall k \in \mathcal{D}_2
\end{equation}
\begin{equation}
\label{eq:potential_change3}
{{u_n}(a_k,{a^*_{{J_k}}}) - {u_n}({a_k},{a_{{J_k}}})}  = 0,\forall k \in \mathcal{D}_3, k \ne n
\end{equation}
 Based on (\ref{eq:potential_change1}) and (\ref{eq:potential_change2}), we have
 \begin{equation}
\label{eq:potential_change4}
  \sum\limits_{k \in \mathcal{D}_1} \big\{{u_k}(a_k,{a^*_{{J_k}}}) - {u_k}({a_k},{a_{{J_k}}})\big\} = |\mathcal{D}_1|
=\sum \limits_{e \in \mathcal{C}_1}  s_n (e,a_{J_{n}})
\end{equation}
 \begin{equation}
\label{eq:potential_change5}
  \sum\limits_{k \in \mathcal{D}_2} \big\{{u_k}(a_k,{a^*_{{J_k}}}) - {u_k}({a_k},{a_{{J_k}}})\big\} = -|\mathcal{D}_2|
=-\sum \limits_{e \in \mathcal{C}_2}  s_n (e,a_{J_{n}})
\end{equation}
Now, combining (\ref{eq:utility_change}), (\ref{eq:potential_change3}), (\ref{eq:potential_change4}) and (\ref{eq:potential_change5}) yields the following equation:
\begin{equation}
\label{eq:equal_change}
 \Phi  (a_n^*,{a_{ - n}}) - \Phi ({a_n},{a_{ - n}})={u_n}(a_n^*,{a_{-n}}) - {u_n}({a_n},{a_{-n}}),
\end{equation}
which  satisfies the definition of EPG, as characterized by  (\ref{eq:EPG_definition}). Thus, the formulated spectrum access game
$\mathcal{F}$ is an EPG, which has at least one pure strategy Nash equilibrium. Furthermore, according to the relationship between
the potential function and the network-centric optimization objective, Theorem \ref{tm:potential_game} is proved.
\end{IEEEproof}

\begin{theorem}
\label{tm:lower_bound}
For any  network topology, the aggregate interference level of all the players at any NE point is  bounded by  $U(a_\text{NE}) \ge  -\frac{\sum\nolimits_{n\in\mathcal{N}} \sum\nolimits_{j \in \mathcal{J}_n} K_n K_j}{M}$.
\end{theorem}
\begin{IEEEproof}
For any pure strategy NE  $a_{\text{NE}}=(a^*_1,\ldots,a^*_N)$, the following inequality holds for each player $n$, $\forall n \in \mathcal{N}$:
\begin{equation}
\label{eq:inequality0}
 {u_n}({a^*_n},a^*_{\mathcal{J}_n}) \ge  {u_n}(\bar a_n,a^*_{\mathcal{J}_n}), \forall \bar a_n \in \mathcal{A}_n, \bar a_n \ne a^*_n,
\end{equation}
 which is obtained according to the definition given in (\ref{eq:NE_definition}). Based on (\ref{eq:inequality0}), it follows that:
\begin{equation}
\label{eq:inequality1}
C^{K_n}_M \times  {u_n}({a^*_n},a^*_{\mathcal{J}_n}) \ge  \sum \limits_{\bar a_n \in \mathcal{A}_n} {u_n}(\bar a_n,a^*_{\mathcal{J}_n}),
\end{equation}
where $C^{K_n}_M$ is the number of $K_n$-combinations  of the channel set $\mathcal{A}_n$ (Note that $|\mathcal{A}_n|=M$).
It is seen that $\sum\nolimits_{\bar a_n \in \mathcal{A}_n} {u_n}(\bar a_n,a^*_{\mathcal{J}_n})$ represents the aggregate
 experienced interference level of player $n$ as if it would access all possible channel profiles simultaneously while the neighboring users still only transmit on their chosen channels. As a result, it can be calculated as follows:
\begin{equation}
\label{eq:inequality2}
\sum\limits_{\bar a_n \in \mathcal{A}_n} {u_n}(\bar a_n,a^*_{\mathcal{J}_n})= -C^{K_n-1}_{M-1}  \sum\limits_{j \in \mathcal{J}_n} K_j,
\end{equation}
where $|\mathcal{J}_n|$ is the number of neighboring users of user $n$. Thus, equation (\ref{eq:inequality1}) can be re-written as:
\begin{equation}
\label{eq:inequality6}
{u_n}({a^*_n},a^*_{\mathcal{J}_n}) \ge  -\frac{C^{K_n-1}_{M-1}}{C^{K_n}_{M}}  \sum\limits_{j \in \mathcal{J}_n} K_j=-\frac{1}{M}\sum\limits_{j \in \mathcal{J}_n} K_nK_j,
\end{equation}
Finally, it follows that:
\begin{equation}
\label{eq:inequality7}
U(a_{NE})=\sum\limits_{n\in\mathcal{N}}{u_n}({a^*_n},a^*_{\mathcal{J}_n}) \ge  -\frac{\sum\limits_{n\in\mathcal{N}} \sum\limits_{j \in \mathcal{J}_n} K_n K_j}{M}
\end{equation}
which proves Theorem \ref{tm:lower_bound}.
\end{IEEEproof}

Theorem \ref{tm:lower_bound} characterizes the achievable interference bound of the formulated spectrum access game. Some further discussions are given below:
\begin{itemize}
  \item If all the players choose only one channel for transmission, i.e., $K_n=1, \forall n \in \mathcal{N}$, we have $U(a_{NE}) \ge  -\frac{\sum\nolimits_{n\in\mathcal{N}} |\mathcal{J}_n|}{M}$.
  \item When the number of available channels increases, the bounded aggregate interference level decreases. In particular, if the number of channels becomes sufficiently large, i.e., $M \rightarrow \infty$, we have $U(a_{NE}) \rightarrow  0$. In this case, the spectrum resources are abundant and mutual interference among the players are completely eliminated. Also, when the network becomes sparse, i.e., decreasing $|\mathcal{J}_n|$, the bounded aggregate interference level also decreases.
\end{itemize}

\vspace{-0.2in}
\subsection{Autonomous best response learning }
As the distributed spectrum access problem now formulated as an exact potential game, the best response (BR) algorithm \cite{Monderer96} can be applied to achieve Nash equilibria of the game. However, there is a limitation of standard BR algorithm:  only one player is randomly selected to update its action in each iteration. However, the convergence speed is very slow when the network becomes dense. To overcome this problem, we exploit the local interference of small cell networks, and propose an autonomous best response learning algorithm, which converges to the NE rapidly.

\begin{figure}[!tb]
\centering
\includegraphics[width=2.1in]{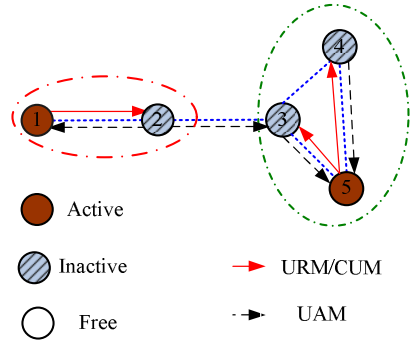}
\caption{The illustrative diagram of the proposed autonomous best response learning algorithm. Using the autonomous contention mechanism, users 1 and 5 can update channel selections simultaneously.}
\label{fig:learning}
\end{figure}

The key idea that multiple users are autonomously selected to update their selections simultaneously. Specifically, due to the local interference among the users, multiple non-neighboring can using the BR rule to update their channel selections \cite{Xu_IEICE_2013}. To achieve this, we assume that there is a common control channel (CCC) available and a 802.11 DCF-like contention mechanism can be applied at each CR user. Specifically, each SAP has three states: \emph{free}, \emph{active}, and \emph{inactive}, as shown in Fig. \ref{fig:learning}, and only the active users have the opportunities to update their channel selections. A brief description of the state transition is as follows:
\begin{itemize}
  \item A free SAP generates a backoff timer according to uniform distribution on an interval, say $[0,\tau_{max}]$ for some fixed parameter $\tau_{max}$. If the backoff timer expires, it becomes
      active. Then, it broadcasts an updating request message (URM).
  \item If  a free SAP hears a URM  before its backoff timer expires, it freezes the timer immediately, enters into the inactive state, and responds with an updating announce message (UAM).
  \item If a free SAP hears a UAM before its backoff timer expires, it also freezes the timer immediately, enters into inactive state, and keep silent until the next period.
      \item When an active SAP receives a UAM  from its neighbors, it updates their channel selection using the BR rule. After the updating, it broadcasts a channel updating message (CUM) to announce its new channel selection, and they becomes free again.
  \item On hearing a CUM message, the inactive users turn to be free again.
\end{itemize}

\begin{figure}[tb!]
\rule{\linewidth}{1pt}
\emph{\textbf{Algorithm 1}:  Autonomous best response algorithm }
\\
\rule{\linewidth}{1pt}
\begin{algorithmic}
\STATE \textbf{1). Initialization:} All the users exchange information (the channel selection) with its neighbors.

\STATE \textbf{2). All SAPs repeatedly perform the following procedure}:
\\
\STATE \quad Based on the information of its neighbors, each SAP $n$ finds the best action selection $n$ as follows:
\begin{equation}
\label{eq:update_rule2}
a^{(b)}_n(i-1)=\arg \mathop {\max }\limits_{a_n \in A_n} u_n(a_n,a_{\mathcal{J}_n}(i-1)),
\end{equation}
where $a_{\mathcal{J}_n}(i-1)$ is the action profiles of its neighboring SAPs in the $(i-1)th$ iteration. That is,  SAP $n$ finds the action $a^{(b)}_n(i-1)$ that maximizes its  utility function given the action profiles of the neighboring SAPs.
\STATE  \quad if $ a^{(b)}_n(i-1)$ is better than the current selection $a_n(i-1)$, SAP $n$ contends for an updating opportunity and updates its channel selection as $a_k(i)= a^{(b)}_k(i-1)$ if the contention is successful; otherwise, it keeps silent.
 \end{algorithmic}
 \rule{\linewidth}{1pt}
\end{figure}

\begin{theorem}
\label{tm:learning_convergence}
The proposed autonomous best response learning algorithm converges to a pure strategy NE point of the formulated spectrum access game $\mathcal{F}$ in finite steps. Therefore, the aggregate interference level in the small cell networks is globally or locally minimized.
\end{theorem}

\begin{IEEEproof}
From the learning procedure, it is seen that each updating user always makes its utility function increasing.
As the updating users are non-neighboring, the potential function of the game, as specified by (\ref{eq:potential}), is  increasing. Since the potential function is up bounded (the maximum value is zero), the learning algorithm will finally converge to a global or local maximum point of the potential function in finite steps. Thus, Theorem \ref{tm:learning_convergence} is proved.
\end{IEEEproof}

Surely, we can achieve the global optimal solutions as the potential function coincides with the objective function of the centralized problem $P1$, using the spatial adaptive play \cite{MAC_Interference1} or B-logit learning \cite{MAC_Interference3}.
However, the convergence speed of the optional algorithms is slow. Therefore, to make it more practically, it should balance the tradeoff between convergence speed and performance, which is the motivation of the proposed autonomous learning algorithm.

\textbf{Remark}. Some discussions on the practical implementations of the autonomous best response learning algorithm are listed below: i) Note that it is assumed that the users are truthful in exchanging information and are obedient in executing the contention mechanism as well as the BR update rule. In essence, the presented game-theoretic solution follows the so-called "engineering agenda" of game theory, i.e., using games as a tool for distributed control \cite{Marden_book}.
ii) only at the initialization phase, each SAP needs to know the current channel selection profiles of neighboring SAPs. In practice, information exchange among neighboring SAPs can be achieved via the backhaul network or the X2 interference. iii) as the algorithm begins to iterate, the users broadcast their new selections in the CUM message, which means that information exchange is not intentionally needed anymore.

\vspace{-0.1in}
\section{Simulation Results and Discussion}

We consider a small cell network deployed in a square region. When there are 20 small cells, the square region is 200m $\times$ 200m. When the number of small cells increases, the square region increases proportionally  to keep the same density. The coverage distance of each small cell is 20m, and the interference distance  is 60m. For presentation, the load of each cell is randomly chosen from a load set $L=\{1,2,3\}$.

\begin{figure}[!tb]
\centering
\includegraphics[width=2.5in]{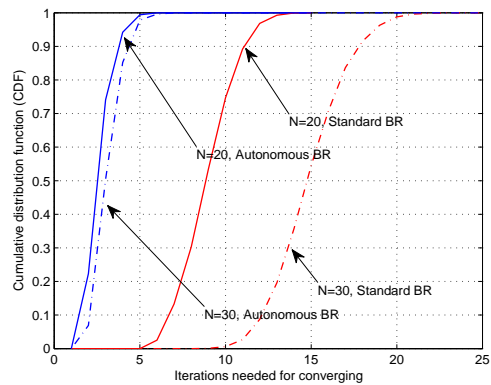}
\caption{The convergence speed comparison between the standard BR and the autonomous BR. (The number of channels is $M=5$) }
\label{fig:convergence}
\end{figure}

To begin with, we compare the convergence speed of the autonomous BR and the standard BR. In the standard BR, only one active user is scheduled to update its action in each iteration, which can be achieved by token or a gateway.
There are five channels available in the network and the companion results of the cumulative distribution function (CDF) of the iterations needed for converging are shown in Fig. \ref{fig:convergence}. The results are obtained by simulating five different network topologies and 1000 independent trials for each network topology. It is noted from the figure that for the same size network, e.g., $N=20$ or $N=30$, the iterations needed for converging of the autonomous BR learning algorithm is significantly decreased. Furthermore, when the network scales up from $N=20$ to $N=30$, the convergence speed of the autonomous BR is slightly decreased while that of the standard BR is largely decreased. The reason is that multiple non-neighboring users can update simultaneously. Thus, the proposed autonomous BR algorithm is especially suitable for large-scale networks.

\begin{figure}[!tb]
 \centering
\includegraphics[width=2.5in]{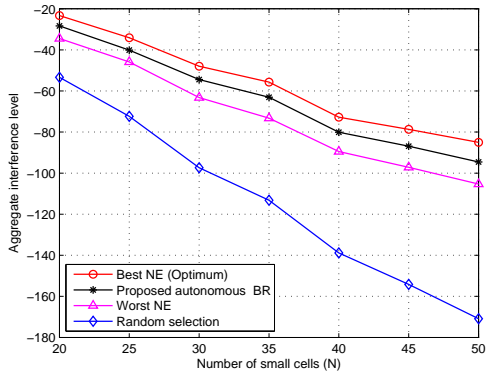}
\caption{The aggregate interference level when varying the number of small cells (The number of channels is $M=5$) }
\label{fig:performance}
\end{figure}

\begin{figure}[!tb]
 \centering
\includegraphics[width=2.5in]{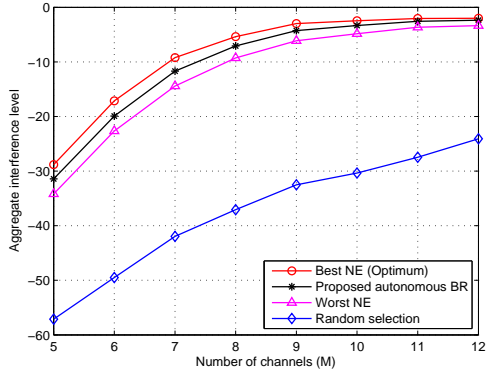}
\caption{The aggregate interference level when varying the number of channels (The number of small cells is $N=20$) }
\label{fig:performance_vs_channel}
\end{figure}

Secondly, the aggregate interference level when varying the number of small cells is shown in Fig. \ref{fig:performance}. The best and worst NE are obtained in a quasi-centralized manner. Specifically, assume there is an omnipotent genie, which knows the cell loads and the interference relation between the SAPs. We run the standard BR learning algorithm 1000 times and then choose the best (worst) result respectively. According to Theorem 1, the best NE also serves as global minimum for the formulated dynamic spectrum access game. It is noted from the figure that as the network scale increases, the aggregate level increases, as can be expected. More importantly, it is noted that the performance of the proposed autonomous best response algorithm is close to the optimum solution. Also, the game-based solution significantly outperforms the random selection strategy. In addition, the  aggregate interference level when varying the number of channels is shown in Fig. \ref{fig:performance_vs_channel}. It is noted that as the number of channels increases, the interference level decreases as can be expected. In particular, as the number of channels is large, e.g., $M>9$, the interference level becomes moderate. Moreover, the performance of the autonomous BR algorithm is close to the optimum.

To summarize, the simulation results show that the proposed game-theoretic converges rapidly and its performance is close to the optimum solution. More importantly, it is scalable when increasing the number of small cells, which means that it is suitable in large-scale networks.


\vspace{-0.1in}
\section{Conclusion}
In this letter, we investigated the problem of self-organizing spectrum access for small cell networks, using a graphical game approach. Compared with existing work, we took the features of different cell loads and local interference relationship into account.  It is proved that the formulated spectrum access game is an exact potential game with the aggregate interference level as the potential function, and Nash equilibrium (NE) of the game corresponds to the global or local optima of the original problem. Also, a lower bound of the  aggregate interference level was rigorously derived. Then, we proposed an autonomous  best response learning algorithm to converge towards NE of the game. It is shown that the proposed learning algorithm converges rapidly and its performance is close to the optimum solution.


\ifCLASSOPTIONcaptionsoff
  \newpage
\fi



%
\bibliographystyle{IEEEtran}
\bibliography{IEEEabrv,reference}

%

%








\end{document}